\documentclass[prb,showpacs,amsmath,amssymb,superscriptaddress,twocolumn,
floatfix]{revtex4}

\usepackage[english]{babel}
\usepackage{graphicx}
\usepackage{times}
\usepackage{units}

\begin{document}

\title{Convergence acceleration and stabilization for dynamical-mean-field-theory
calculations}

\author{Rok \v{Z}itko}
\affiliation{Jo\v{z}ef Stefan Institute, Jamova 39, SI-1000 Ljubljana, Slovenia}

\date{\today}


\begin{abstract}
The convergence to the self-consistency in the dynamical-mean-field-theory
(DMFT) calculations for models of correlated electron systems can
be significantly accelerated by using an appropriate mixing of hybridization
functions which are used as the input to the impurity solver. It is shown
that the techniques and the past experience with the mixing of input charge
densities in the density-functional-theory (DFT) calculations are also
effective in DMFT. As an example, the increase of the computational
requirements near the Mott metal-insulator transition in the Hubbard model
due to critical slowing down can be strongly reduced by using the modified
Broyden's method to numerically solve the non-linear self-consistency
equation. Speed-up factors as high as 3 were observed in practical
calculations even for this relatively well behaved problem. Furthermore, the
convergence can be achieved in difficult cases where simple linear mixing is
either not effective or even leads to divergence. Unstable and metastable
solutions can also be obtained.
We also determine the linear response of the system with respect to the
variations of the hybridization function, which is related to the
propagation of the information between the different energy scales during
the iteration.
\end{abstract}

\maketitle

\newcommand{\korr}[1]{\langle\langle #1 \rangle\rangle}
\renewcommand{\Im}{\mathrm{Im}}
\renewcommand{\Re}{\mathrm{Re}}
\newcommand{\vc}[1]{\mathbf{#1}}

\section{Introduction}

In many transition-metal, lanthanide, and actinide compounds the
almost-localized $d$ and $f$ orbitals are partially filled and local
magnetic moments are formed at low temperatures \cite{hewson, imada1998}.
The competition between itinerancy and local electron-electron correlation
effects gives rise to complex phase diagrams with different magnetic,
charge-ordered, and superconducting phases \cite{dagotto1994, stewart1984,
hewson, imada1998, salamon2001, lee2006}. Simplified tight-binding models
with short-range Coulomb interaction terms are commonly used to study
strong-correlation effects in such systems. In the paradigmatic Hubbard
model \cite{anderson1959a, hubbard1963, kanamori1963, gutzwiller1963}, the
problem is reduced to a single-orbital description with purely on-site
Coulomb repulsion. Hubbard-like models are thought to correctly describe
certain aspects of the itinerant ferromagnetism, the metal-insulator
transitions \cite{brinkman1970, imada1998}, and the high-temperature
superconductivity \cite{anderson1987, dagotto1994}. Despite intensive
research, the properties of the Hubbard model are not yet fully established.
In the limit of infinite dimensions or high lattice connectivity the problem
can be solved by the dynamical mean-field theory (DMFT) \cite{metzner1989,
mullerhartmann1989, georges1992, jarrell1992dmft, rozenberg1992, zhang1993,
sakai1994, pruschke1995, maier2005, georges1996}. In this limit, the
self-energy becomes purely local and the bulk problem of correlated
electrons maps exactly onto a quantum impurity model with a
self-consistently-defined non-interacting bath of conduction electrons. In
the DMFT, the spatial correlations are described in a mean-field way,
however the local quantum fluctuations are taken into account exactly; as
long as the effect under study is driven by local physics, the results of
DMFT calculations are a good approximation for the properties of real
(finite-dimensional) materials.

In spite of the significant simplification of the full problem within the
DMFT, the solution of the effective quantum impurity problem is still
challenging: it is by far the most computationally demanding part of the
calculations. Among several impurity solvers in common use, the numerical
renormalization group (NRG) \cite{wilson1975, krishna1980a, sakai1989,
costi1994, bulla1998, bulla1999, pruschke2000, bulla2008} is distinguished
by its applicability to study the regime of very low temperatures directly
in the thermodynamic limit. The convergence-acceleration approach proposed
in the following is clearly applicable to any impurity solver that may be
used to solve the DMFT problem, however the discussion, the implementation
details, and the test results are given for the NRG. Nevertheless, the
technique can easily be adapted for other solvers in a straight-forward
manner.

The input to a NRG calculation is the hybridization function
$\Gamma(\omega)$ which contains information about the density of states of
the effective medium into which the impurity is embedded, while the output
is an impurity spectral function $A(\omega)$, which is then used to compute
the local lattice spectral function $\rho(\omega)$. The self-consistency is
achieved when the two become equal, i.e. $A(\omega)=\rho(\omega)$ within
chosen accuracy, otherwise $\rho(\omega)$ is used to compute new
hybridization function for the next DMFT iteration. In order to ensure the
convergence, in some situations $\Gamma(\omega)$ from two consecutive
iterations are linearly mixed to obtain the hybridization function which is
used as the input to the NRG.
A similar situation is well known in the field of quantum chemistry and
electronic-structure calculations, in particular in the
density-functional-theory (DFT) \cite{hohenberg1964, kohn1965, martin} where
the quantity to be mixed is the charge density in space, $n(\vc{r})$. In
difficult cases (metallic surfaces, heterostructures, impurities in metals,
systems near magnetic instabilities, etc.), the simple linear mixing
procedure converges too slowly (or not at all), thus more sophisticated
mixing approaches were devised \cite{dederichs1983}. In these schemes, a
system of nonlinear equations is solved iteratively by quasi-Newton-Raphson
procedures or similar methods \cite{broyden1965, anderson1965, pulay1980,
dederichs1983, vanderbilt1984, srivastava1984, johnson1988, eyert1996,
kresse1996, kresse1996comp, bowler2000, baran2008, marks2008, anglade2008}.
The solution of the system is equivalent to the Kohn-Sham variational
principle \cite{bendt1982}. 
Such ``advanced mixing'' techniques are implemented with various degrees of
sophistication in all DFT packages. They are stable, easy to implement and
use, and they often have very high convergence rate.

In this paper it is shown that the techniques and the past experience with
the advanced mixing schemes in DFT calculations can also be applied to DMFT
calculations. The advanced mixing greatly accelerates the convergence in
many cases, for example near the Mott metal-insulator transition, where the
iteration converges very slowly due to critical slowing-down. It also
ensures the convergence to unstable and metastable solutions, hence it can
be applied to situations with multiple coexisting solutions.

This work is structured as follows. In Sec.~\ref{sec2} the DMFT
self-consistency constraint is formulated as a sufficient condition in the
form of a system of equations. In Sec.~\ref{sec3} the modified Broyden's
iterative method for solving systems of nonlinear equations is briefly
described, focusing on the implementation with low storage requirements
which is more suitable for large-scale problems \cite{johnson1988}. In
Sec.~\ref{sec4} it is shown how the Broyden solver is incorporated into the
DMFT loop and some further implementation details are given. In
Sec.~\ref{sec5} the convergence properties of the linear and advanced mixing
schemes are compared on the example of the Hubbard model for increasing
electron-electron repulsion $U$. In Sec.~\ref{sec6} the Hubbard model in
external magnetic field is considered; in this case, the simple mixing is
not always successful and the use of Broyden's method was found to be
essential to obtain rapid convergence. In Sec.~\ref{sec7} we study the
response of the Hubbard model with respect to small variations of the
hybridization function; this response function is the equivalent of the
Jacobian matrix of the system of self-consistency equations and describes
the propagation of the information between various energy scales during the
DMFT iteration. Finally, in Sec.~\ref{sec8} some examples of the
stabilization of otherwise unstable fixed-point solutions are discussed.

\section{DMFT self-consistency constraint as a system of non-linear
equations} \label{sec2}

The single-orbital Hubbard model \cite{anderson1959a, hubbard1963} for
electrons on a $d$-dimensional lattice
\begin{equation}
H = \sum_{\langle ij \rangle,\sigma} t_{i,j}
c^\dag_{i,\sigma} c_{j,\sigma} - \sum_{i,\sigma} \mu_\sigma n_{i,\sigma} 
+ U \sum_i n_{i,\uparrow} n_{i,\downarrow}
\end{equation}
(with $n_{i,\sigma} = c^\dag_{i,\sigma} c_{i,\sigma}$ and
$\mu_\sigma=\mu-(\sigma/2) g \mu_B B$) maps in the $d\to\infty$ limit 
\cite{georges1992, jarrell1992, georges1996} onto the single-impurity
Anderson model \cite{anderson1961}
\begin{equation}
\begin{split}
H_\mathrm{SIAM} &= \epsilon_{d,\sigma} n + U n_\uparrow n_\downarrow +
\sum_{k,\sigma} \left( V_{k,\sigma} c_{k,\sigma}^\dag d_\sigma + 
\text{H.c.} \right) \\ 
&+ \sum_{k,\sigma} \epsilon_{k,\sigma}
c^\dag_{k,\sigma} c_{k,\sigma}
\end{split}
\end{equation}
with $n_\sigma = d^\dag_\sigma d_\sigma$ and $n=n_\uparrow+n_\downarrow$.
The hybridization function $\Gamma_\sigma(\omega)=\sum_k |V_{k,\sigma}|^2
\delta(\omega-\epsilon_{k,\sigma})$ contains full information about the
coupling between the impurity and the effective non-interacting medium. From
the calculated impurity spectral function 
\begin{equation} 
A_\sigma(\omega)= -\frac{1}{\pi} \Im \left[ G_\sigma(\omega+i \delta) \right],
\end{equation}
where $G_\sigma(z)=\korr{d_\sigma ; d^\dag_\sigma}_z$ is the impurity
Green's function, one can extract the interaction self-energy
$\Sigma_\sigma(\omega)$ defined through
\begin{equation}
G_\sigma(\omega) = \frac{1}{\omega-\epsilon_{d,\sigma} 
+ \Delta_\sigma(\omega) - \Sigma_\sigma(\omega)},
\end{equation}
where $\Im \Delta_\sigma(\omega) = \Gamma_\sigma(\omega)$ and the real part
of $\Delta_\sigma(\omega)$ can be obtained via the Kramers-Kronig
transformation. In practice, the self-energy can be calculated more
reliably and accurately as the ratio of two correlation functions
\cite{bulla1998}: the generalized Green's function $F_\sigma(z)=
\korr{d_\sigma n_{\bar{\sigma}} ; d^\dag_\sigma}_z$ over the Green's
function $G_\sigma(z)$, i.e. $\Sigma_\sigma(\omega) = U F_\sigma(\omega) /
G_\sigma(\omega)$. The local lattice Green's function is
\begin{align}
G_{\mathrm{loc},\sigma}(\omega) &= \frac{1}{N} \sum_k G_{k,\sigma}(\omega) \\
&= \frac{1}{N} \sum_k
\frac{1}{\left[\omega+\mu_\sigma-\Sigma_\sigma(\omega)\right]-\epsilon_{k}} \\
&= \int \frac{\rho_0(\epsilon) d\epsilon}{\left[
\omega+\mu_\sigma-\Sigma_\sigma(\omega) \right]-\epsilon},
\end{align}
where $\rho_0(\epsilon)$ is the density of states (DOS) in the
noninteracting model. The local lattice spectral function is then
\begin{equation}
\rho_\sigma(\omega) = -\frac{1}{\pi} \Im \left[
G_{\mathrm{loc},\sigma}(\omega+i\delta) \right].
\end{equation}
The self-consistency condition \cite{georges1996} relates the local lattice
Green's function $G_{\mathrm{loc},\sigma}$ and the hybridization function
$\Gamma_\sigma$ as
\begin{align}
\label{Gamma}
\Gamma_\sigma(\omega) &= -\Im \left[ 
\omega-\mathcal{G}_{0,\sigma}^{-1}(\omega) \right], \\
\mathcal{G}_{0,\sigma}^{-1}(\omega) &= G_{\mathrm{loc},\sigma}^{-1} +
\Sigma_\sigma(\omega).
\end{align}

One DMFT cycle (which involves the numerical solution of $H_\mathrm{SIAM}$,
the calculation of $G_{\mathrm{loc},\sigma}$, and the determination of the
new hybridization function $\Gamma_\sigma$ via Eq.~\eqref{Gamma}) can be
considered as a functional of the input hybridization function, i.e.
\begin{equation}
\Gamma^\mathrm{new}_\sigma = \Gamma^\mathrm{new}_\sigma \left\{
\Gamma^\mathrm{old}(\omega) \right\}
\end{equation}
If the self-consistency has been established, the hybridization function
is invariant (fixed point):
\begin{equation}
\label{fixedpoint}
\Gamma^\mathrm{new}_\sigma  \left\{
\Gamma^\mathrm{old}_\sigma(\omega) \right\} = \Gamma^\mathrm{old}_\sigma.
\end{equation}
Defining a mapping $F$ as the difference
\begin{equation}
\label{eq10}
F(\Gamma_\sigma) = \Gamma^\mathrm{new}_\sigma \left\{ \Gamma_\sigma \right\}
- \Gamma_\sigma,
\end{equation}
the approach to the self-consistency clearly corresponds to solving the
system of equations 
\begin{equation}
\label{F}
F(\Gamma_\sigma)=0,
\end{equation}
while a single DMFT step corresponds to applying $F$ once to the
hybridization function. Any solution of the equation \eqref{F} is a possible
physical state of the system since it satisfies the self-consistency
condition, albeit it is not necessarily the ground state: solutions
corresponding to unstable and metastable states can also be found (see
Sec.~\ref{sec8}). It should be noted that Eq.~\eqref{F} is highly
non-linear.

The usual DMFT iteration with no mixing corresponds to solving Eq.~\eqref{F}
by a direct iteration, which often works since the mapping $F$ behaves as a
contraction in the vicinity of the solution (and often even far away from
it). When $F$ is not a contraction, however, this procedure will tend to
diverge and more care is required. Usually it is sufficient to take an
average of two hybridization functions (the current and the previous one):
\begin{equation}
\Gamma^{\mathrm{input},(m)} = \alpha \Gamma^\mathrm{new,(m)} + (1-\alpha)
\Gamma^\mathrm{input,(m-1)},
\end{equation}
where $\alpha \in [0:1]$ is the mixing parameter. 
It should be remarked that this is fully analogous to simple charge mixing
in the density-functional theory, where the charge density from the previous
iteration $n^\mathrm{old}(\vc{r})$ is admixed to the current one
$n^\mathrm{new}(\vc{r})$ as the true input to the next DFT iteration.
Unfortunately, there are situations where this simple linear mixing approach
fails even for small values of $\alpha$. Furthermore, for very small
$\alpha$ the approach to the self-consistency becomes prohibitively slow. In
such situations, more sophisticated mixing approaches are required. In DFT,
Broyden's method is commonly used.

\section{Broyden's method}
\label{sec3}

Let $\vc{V}$ be an $N$-dimensional vector and $F$ a mapping; the goal is to
solve the system of equations $F(\vc{V})=0$. The quasi-Newton-Raphson
methods are iterative techniques in which the new approximation is
given by
\begin{equation}
\vc{V}^{(m+1)} = \vc{V}^{(m)} - \left[ J^{(m)} \right]^{-1} \vc{F}^{(m)},
\end{equation}
where $J^{(m)}$ is the Jacobian of the system at point $\vc{V}^{(m)}$ and
$\vc{F}^{(m)}=F\left( \vc{V}^{(m)} \right)$. The true Jacobian is unknown;
a simple approximation is used for the initial Jacobian, for example
a constant diagonal matrix
\begin{equation}
\label{init}
J^{(1)} = -\frac{1}{\alpha} \mathbf{1},
\end{equation}
which corresponds to simple linear mixing with a mixing parameter $\alpha
\in [0:1]$. The approximation is then improved by performing rank-1 updates
as the iteration proceeds. It is more efficient to update directly the
inverse of the Jacobian $B^{(m)} = -\left[ J^{(m)} \right]^{-1}$ as
\cite{broyden1965, srivastava1984}
\begin{equation}
B^{(m+1)} = B^{(m)} + \left( \Delta \vc{V}^{(m)}-B^{(m)} \Delta \vc{F}^{(m)}
\right) \otimes \Delta\vc{F}^{(m)},
\end{equation}
where 
\begin{align}
\Delta \vc{V}^{(m)} &= \frac{ \vc{V}^{(m+1)} - \vc{V}^{(m)} }
{ \left|
\vc{F}^{(m+1)} - \vc{F}^{(m)} 
\right| }, \\
\Delta \vc{F}^{(m)} &= \frac{ \vc{F}^{(m+1)} - \vc{F}^{(m)} }
{ \left|
\vc{F}^{(m+1)} - \vc{F}^{(m)} 
\right| }.
\end{align}
Vanderbilt and Louie have proposed a modified version of Broyden's method in
which the information from all previous iterations is incorporated when the
current Jacobian is updated; this approach has better convergence properties
and the Jacobian converges to the true Jacobian, which is not the case in
the original Broyden's method which only uses the information from the most
recent iteration to perform the update \cite{vanderbilt1984}. Srivastava has
simplified the computational scheme so that only the input vectors
$\vc{V}^{(m)}$ and output vectors $\vc{F}^{(m)}$ need to be stored, rather
than the complete Jacobian matrix \cite{srivastava1984}. Johnson combined
the advantages of both schemes without any increase in complexity
\cite{johnson1988}. The final expressions for this modified
Broyden's method are as follows:
\begin{equation}
\label{update}
\vc{V}^{(m+1)} = \vc{V}^{(m)} + \alpha \vc{F}^{(m)}
- \sum_{n=1}^{m-1} \sum_{k=1}^{m-1} w_n w_k c^{(m)}_k \beta_{k,n}^{(m)} \vc{U}^{(n)}
\end{equation}
with
\begin{align}
c_{k}^{(m)} &= \left( \Delta \vc{F}^{(k)} \right)^\dag \vc{F}^{(m)}, \\
\vc{U}^{(n)} &= \alpha \Delta \vc{F}^{(n)} + \Delta \vc{V}^{(n)},
\end{align}
and $(m-1) \times (m-1)$ dimensional matrices
\begin{align}
\beta^{(m)}_{k,n} &= \left[ \left( w_0^2 \mathbf{1} +
A^{(m)} \right)^{-1}
\right]_{k,n}, \\
A^{(m)}_{k,n} &= w_k w_n (\Delta \vc{F}^{(n)})^\dag \Delta \vc{F}^{(k)}.
\end{align}
Here $\mathbf{1}$ is a $(m-1)\times(m-1)$ dimensional identity matrix. The
first two terms in Eq.~\eqref{update} correspond to simple linear mixing
with parameter $\alpha$, as described above, while the final term is a
correction which takes into account the updates to the initial Jacobian.

The weights $w_n$ ($n=1,2,\ldots$) are usually chosen to be equal to 1,
while $w_0=0.01$ \cite{johnson1988, baran2008}. For a suitable choice
of weights, the modified Broyden's method becomes equivalent
\cite{kresse1996comp, eyert1996} to Pulay mixing scheme \cite{pulay1980}
or Anderson mixing scheme \cite{anderson1965}.

The algorithm can be simply modified to use only a finite number of previous
iterations to update the vector. This may be advantageous when the initial
approximation for the vector is not very good. Alternatively, the Broyden
mixing can be fully restarted after a given number of iterations.

\section{Incorporation of the Broyden solver into the DMFT loop}
\label{sec4}

In the proposed convergence acceleration scheme for DMFT, the modified
Broyden's method is used to refine the hybridization function
$\Gamma(\omega)$ which is used as the input to the impurity solver. It
should be remarked that this is not the only possibility: alternatively, one
could also mix the self-energy $\Sigma(\omega)$. The choice depends somewhat
on the problem and for numerical reasons one should in extreme cases 
preferably choose the quantity which is smoother as a function of the energy
(for example, near the Mott transition on the metallic side the self-energy
features sharp peaks while the hybridization function is rather smooth,
whereas in the antiferromagnetic phase with small $U$ the hybridization
function contains sharp inverse-square-root singularities while the
self-energy is relatively smooth). In general, however, the two approaches
are expected to be nearly
equivalent. %

\begin{figure}[htbp]
\includegraphics[width=8cm,clip]{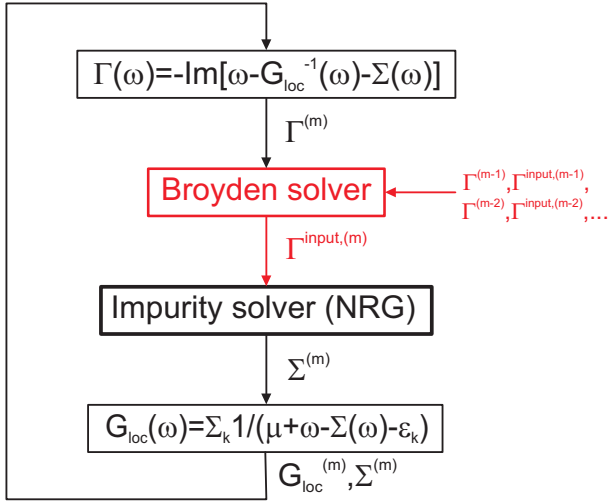}
\caption{(Color online) The DMFT loop using the numerical renormalization
group (NRG) as the impurity solver. The Broyden solver is incorporated in 
the loop as a correction step which modifies the input hybridization
function in order to accelerate the convergence to the self-consistency. The
new elements in the loop are shown in gray (red online).}
\label{fig0}
\end{figure}

The Broyden solver is called just before the NRG, see Fig.~\ref{fig0}. In a
sense, the Broyden solver is effectively driving the DMFT loop in order to
solve the equation 
\begin{equation}
F\left\{ \Gamma^{\mathrm{input},(m)} \right\} =
\Gamma^{(m+1)}\left\{  \Gamma^{\mathrm{input},(m)} \right\}
-
\Gamma^{\mathrm{input},(m)} = 0,
\end{equation}
see also Eqs.~\eqref{eq10} and \eqref{F}. One cycle of the loop thus
corresponds to applying once the mapping $F$ to the hybridization function.

The vector $\vc{V}^{(m)}$ corresponds to a discretized representation of the
continuous function $\Gamma(\omega)$. In the calculations presented in this
work, we used a geometric sequence of points $\omega_{\pm,n} = \pm
\Omega_\mathrm{max} l^n$ with $\Omega_\mathrm{max}$ that exceeds the
bandwidth of the noninteracting band by a factor of order 10 and $l=1.01$;
the dimension of the vectors was $N=3982$ (and twice as large for
spin-dependent problems). The same grid is used to sample all other
functions, in particular the impurity spectral function $A(\omega)$ 
and the self-energy $\Sigma(\omega)$. The Jacobian matrix in the Broyden
solver was typically initialized with $\alpha=1$. The weights were chosen as
$w_0=0.01$ and $w_n=1$ for $n\geq1$; setting $w_0$ to zero was found to have
little effect. It has been suggested that the weights $w_n$ be chosen as
$w_n = \langle \vc{F}^{(m)} | \vc{F}^{(m)} \rangle^{-1/2}$, i.e. as the
inverse root-mean-square difference of the function \cite{johnson1988}.
Numerical tests have shown that the improvement is only minor, if at all
existing.

During the initial steps it sometimes occurs that the resulting
$\Gamma^\mathrm{input}(\omega)$ is not positive for all $\omega$ as the
solver is overcompensating for the deviations. In such cases, the function
was simply clipped to positive values. When the error vectors $\Delta
\vc{F}$ become smaller as the iteration proceeds, this is no longer a
problem. The clipping performed during the initial iterations does not
affect the final result. An alternative solution would be to revert to
simple linear mixing in such instances. Yet another possible approach to
enforce positivity of $\Gamma$ would consist of working with $\ln \Gamma$
instead. Unfortunately, this method was found to slow down the convergence
significantly.

It is necessary to store both $\Gamma^{(m)}(\omega)$ and
$\Gamma^{\mathrm{input},(m)}(\omega)$ for all $N_\mathrm{steps}$ DMFT steps,
thus the additional storage requirements are of the order of $N \times
N_\mathrm{steps}$, which is not likely to pose difficulty.

In calculations with fixed occupancy (rather than fixed chemical potential)
it is important to store as an additional component of the vector $\vc{V}$
also the chemical potential $\mu$ that is being tuned. (This is actually
true in general: all parameters varied in the iteration should appear in the
Broyden process, so that the output of a single iteration is a smooth and
uniquely defined function of the input vector $\vc{V}$ alone
\cite{baran2008}.) In fact, the tuning of the parameter $\mu$ can be integrated
in the Broyden solver with much fruition.

In the NRG calculations performed for testing the method and presented in
the following, the $z$-averaging \cite{frota1986, oliveira1994, campo2005}
over $N_z=8$ values of the twist parameter was used in combination with an
improved discretization scheme based on solving a differential equation to
obtain the discretization coefficients \cite{resolution, odesolv}. The
discretization parameter was $\Lambda=2$, the truncation cutoff was set to
$E_\mathrm{cutoff}=10\omega_N$ (but no less than 500 and no more than 10000
states were used) and care was taken to truncate in a ``gap'' between
clustered excitation levels. Spectral functions were computed using the
density-matrix approach \cite{hofstetter2000} and the self-energy trick
\cite{bulla1998}.  Spectral information was extracted from both even and odd
NRG iterations with a window parameter $p=2.3$ \cite{resolution}. The
broadening procedure from Ref.~\onlinecite{weichselbaum2007} with
$\alpha=0.1$ was used. The choice of NRG parameters appears to be important
for the convergence: high-quality (smooth) results tend to be beneficial for
the rate of convergence, while ``rough'' calculations sometimes lead to a
stagnation of the convergence and oscillatory behavior. This is related to
the assumption of differentiability of the mapping $F$. For the same reason,
the calculations performed with larger broadening parameter will converge
faster than high-energy-resolution calculations with much smaller broadening
parameter. This is especially true when the hybridization function contains
sharp features.

The DMFT loop is terminated when two consecutive impurity spectral functions
$A(\omega)$ differ by no more than some chosen value:
\begin{equation}
\int |A^{(m)}(\omega)-A^{(m-1)}(\omega)| d\omega \leq \lambda.
\end{equation}
In practice it is found that this convergence test is more stringent when
compared to an equivalent test for consecutive local lattice spectral
functions $\rho(\omega)$, while comparing $A(\omega)$ and $\rho(\omega)$ at
the same iteration gives absolute integrated errors somewhere between these
two convergence tests. A typical convergence limit is $\lambda=10^{-6}$.

The stability of the converged solution can be tested by performing a few
further DMFT iterations with the Broyden mixing turned off. From the
solutions one can extract the dominant eigenvalue and eigenvector of the
mapping $F$. This information is instrumental in assessing the physical
stability of the solution and to determine the type of eventual instability.
We return to these considerations in Sec.~\ref{sec8}.

\section{Acceleration of the convergence}
\label{sec5}

\begin{figure}[htbp]
\includegraphics[width=8cm,clip]{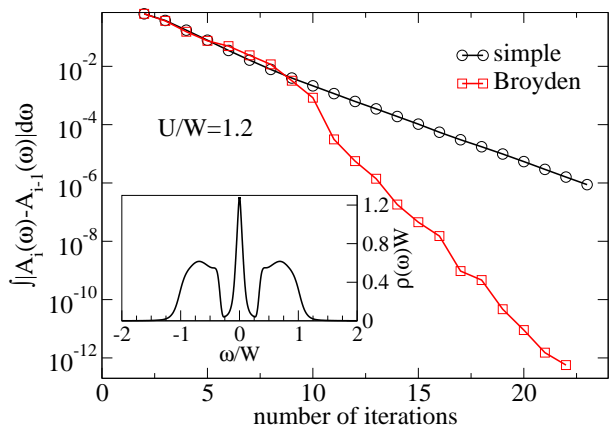}
\caption{(Color online) Comparison of the convergence of the impurity
spectral function $A(\omega)$ in a calculation for the Hubbard model defined
on the Bethe lattice in the paramagnetic phase at half-filling. 
We compare the simple mixing
(here $\alpha=1$, i.e. the output from the previous iteration is used
directly as the input for the new iteration, so this is actually direct
iteration rather than mixing) and Broyden's mixing. The inset shows the
converged density of states.}
\label{fig1}
\end{figure}

The acceleration of the convergence of the DMFT loop towards
self-consistency was explored on the well-studied case of the Hubbard model
at half-filling ($\mu=0$) in the paramagnetic regime \cite{georges1992,
jarrell1992, rozenberg1992, zhang1993, sakai1994, moeller1995, georges1996,
bulla1998, bulla1999, bulla2008}. We study the Hubbard model on the Bethe
lattice with infinite coordination number where 
\begin{equation}
\rho_0(\epsilon)=\frac{4}{\pi W} \sqrt{1-(2\epsilon/W)^2}.
\end{equation}
Here $W$ is the width of the non-interacting conduction band. As the
electron-electron repulsion $U$ is increased, the characteristic three-peak
structure emerges: two Hubbard bands and a quasiparticle peak at the Fermi
level. As $U$ approaches a critical value of $U_\mathrm{c}/W \approx 1.46$,
the quasiparticle peak becomes increasingly narrow and disappears
\cite{moeller1995, georges1996, bulla1998, bulla1999}. Recent
high-energy-resolution calculations have confirmed that the Hubbard bands
have inner structure, in particular a peak at the inner edges
\cite{karski2005, karski2008, resolution}; this structure can be observed,
for example, in the inset in Fig.~\ref{fig1}.

\begin{figure}[htbp]
\includegraphics[width=8cm,clip]{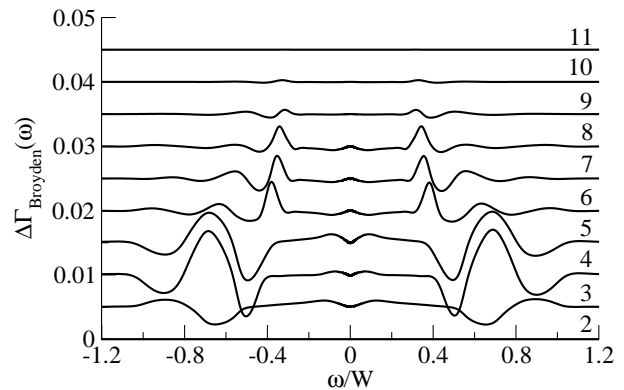}
\caption{
The Broyden corrections for consecutive iterations; the curves are offset
for clarity. The parameters are as in Fig.~\ref{fig1}. }
\label{fig3}
\end{figure}

In Fig.~\ref{fig1} we compare the approach to the self-consistency for the
Hubbard model at fixed $U/W=1.2$. The initial approximation for the local
spectral function was the non-interacting DOS
$\rho_0(\omega)$. Initially, both approaches are equivalent, since the
starting approximation for the Jacobian is a diagonal matrix which
corresponds to simple mixing. Since $\rho_0(\omega)$ is a rather crude
approximation to the real density of states, Broyden's method is not
expected to work much better than simple mixing for the first few steps;
indeed, the errors are found to be even slightly higher. As can be seen in
Fig.~\ref{fig3}, the non-linear Broyden corrections are initially especially
large in the region of the emerging Hubbard bands, while at later iterations
the most important contributions are to the inner-edge peaks in the Hubbard
bands. Starting with iteration 9, the approximation to the self-consistent
hybridization function becomes quite adequate, the updates to the Jacobian
correspond to accurate refinements and the convergence accelerates
significantly.
Both methods converge linearly, however the rate of convergence is much
faster with Broyden's method. Superlinear convergence was never observed in
practice. The linear rate of convergence in the example shown in
Fig.~\ref{fig1} was $\mu \approx 0.15$. When required, extremely good
accuracy of the solution can thus be obtained with essentially no additional
computational effort as compared to the direct iteration.

\begin{figure}[htbp]
\includegraphics[width=8cm,clip]{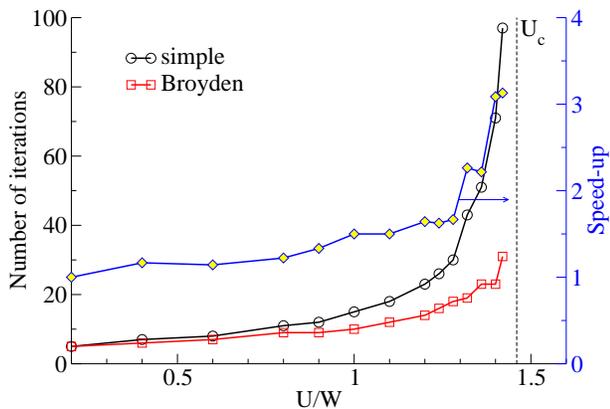}
\caption{(Color online) Comparison of the convergence as a function of the
interaction strength $U/W$. The convergence is defined to occur when two
consecutive solutions for the density of states differ by no more than
$\lambda=10^{-6}$ (integrated absolute value of the difference). The
vertical dashed line corresponds to the point of the Mott metal-insulator
transition at $U_\mathrm{c}/W=1.46$\,.}
\label{fig2}
\end{figure}

We also determined the speed-up due to using the modified Broyden's method
as a function of the interaction strength $U$, Fig.~\ref{fig2}. As the Mott
metal-insulator transition is approached from below, the convergence becomes
more difficult to achieve, which can be assigned to critical slowing down in
the vicinity of quantum phase transitions \cite{bulla1999, rozenberg1999,
joo2001}. Both approaches are affected by this difficulty, however it is
found that the relative speed-up in Broyden's method is an increasing
function of $U$; for the range of parameter $U$ considered in this work, the
speed-up was up to a factor of 3 and it presumably increases even further
for $U \to U_\mathrm{c}$. It should be remarked that in these calculations
it was possible to use $\alpha=1$ (in other words, the direct DMFT iteration
converges without any mixing), which is the most favorable
situation. In problems where mixing with small $\alpha$ is necessary, the
speed-up factor is expected to be much higher.

By performing calculation in the close vicinity of the Mott transition on
the metallic side, we obtained an improved estimate of the critical value of
$U$:
\begin{equation}
U_{\mathrm{c}}/W = 1.459\,.
\end{equation}
It agrees very well with previous NRG calculations, where $U_\mathrm{c}/W =
1.47$ was established \cite{bulla1999}, and even better with the value
obtained using projective self-consistent approach, $U_\mathrm{c}/W=1.46$
\cite{moeller1995, georges1996}.

\section{Hubbard model in the magnetic field}
\label{sec6}

The Hubbard model in a strong magnetic field \cite{laloux1994, bauer2007fm}
has a metamagnetic response in a certain parameter regime: the magnetic
susceptibility increases with the field strength \cite{vollhardt1984,
spalek1990, laloux1994, georges1996, bauer2009fm}. The metamagnetic response
is due to electron-electron interactions and, for sufficiently large $U$, it
is driven mostly by field-induced quasiparticle mass enhancement (i.e.
field-induced localization), however quasiparticle interactions also play a
role \cite{bauer2009fm}.

We consider the Hubbard model at half-filling and at zero temperature in a
magnetic field. This problem is interesting for several reasons: 1) it is
found that a DMFT iteration with simple mixing (and taking the
noninteracting DOS as an initial approximation) does not always converge in
the presence of the magnetic field; 2) the structure at the inner-edge of
the Hubbard band might be of magnetic origin, thus it can have non-trivial
behavior in a finite magnetic field \cite{karski2005, karski2008,
resolution}; 3) the behavior near the threshold to full polarization is not
fully understood due to numerical difficulties in the transition regime
\cite{bauer2009fm}.

\begin{figure}[htbp]
\includegraphics[width=8cm,clip]{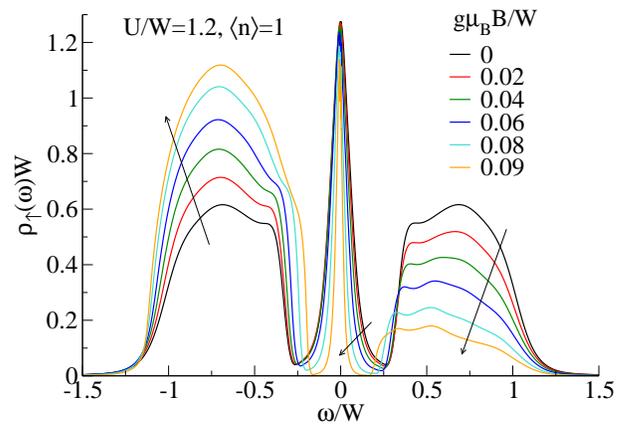}
\caption{(Color online) Majority-spin spectral functions for the half-filled
Hubbard model in a magnetic field. The arrows show the direction of the
increasing magnetic field.%
}
\label{fig5}
\end{figure}

The calculated spectral functions are presented in Fig.~\ref{fig5}. The
results agree with those shown in Ref.~\onlinecite{bauer2009fm}, however the
energy resolution in our approach is sufficiently higher so that the inner
structure in the Hubbard bands may be resolved \cite{resolution}. As already
established, when the magnetic field is increased the quasiparticle peak
shifts away from the Fermi level and it narrows down, and the spectral
weight is gradually transferred to the lower Hubbard band of the majority
spin \cite{bauer2009fm}. With improved resolution, we can now also observe
that the internal structure of the Hubbard bands changes significantly with
increasing field. When the magnetic field is increased past a
transition value $B_{\mathrm{c}}$, a field-induced metal-insulator
transition is induced \cite{bauer2009fm}. 

For a system in the metallic regime, the
Broyden's method converged rapidly, even when the non-interacting DOS was
taken as the initial approximation, while linear mixing usually led to
oscillatory behavior. 
As in the Mott metal-insulator transition, the number of necessary
iterations increases as the transition point is approached. On the
insulating side, the convergence to the fully polarized solution was rapid
for large fields, however the calculations in the vicinity of the transition
point were more difficult and it was necessary to initialize the problem
with the fully-polarized insulating spectral functions to ensure the
convergence. The difficulties appear to stem from the fact that the
non-interacting DOS for the Bethe lattice has square-root singularities at
the band edges, while the Broyden method is premised on the
differentiability of the mapping $F$.

The inner structure in the Hubbard bands remains present even in the
presence of the magnetic field, see Fig.~\ref{fig5}. With increasing field,
the lower Hubbard band of the majority-spin electrons becomes increasingly
featureless and the inner-edge peak tends to disappear as we approach the
transition to the insulating phase. The upper Hubbard band, however, appears
to become more structured and distinctively asymmetric. Even at low fields
there is some hint of further weak peaks within this band, which become more
pronounced in the vicinity of the transition. In this regime, the electrons
are already strongly polarized, thus majority-spin electrons in the upper
Hubbard band cannot easily propagate since they reside on doubly-occupied
sites surrounded predominantly by a ferromagnetic background, thus their
motion is strongly hindered by the Pauli exclusion principle and they become
increasingly localized. This is to be contrasted with the holes in the lower
Hubbard band which can easily propagate and do not feel the strong
electron-electron
interactions. %

\section{Response with respect to the variation of the input hybridization
function} \label{sec7}

Finding a good initial approximation to the Jacobian is not trivial,
therefore a simple diagonal constant matrix is typically used, as in 
Eq.~\eqref{init}. In band-structure calculations, the Jacobian is related to
the dielectric tensor \cite{bendt1982, dederichs1983, srivastava1984}, which
makes it possible to devise an improved initial approximation for the
Jacobian based on the Thomas-Fermi screening theory (this procedure is
called ``preconditioning'') \cite{dederichs1983, srivastava1984, kresse1996,
anglade2008}. The Jacobian for the hybridization function in the DMFT loop
is not related to some well-understood physical quantity in a simple way
(see, however, the discussion of the Landau-Ginzburg functional
$F_\mathrm{LG}$ of the hybridization function discussed in
Ref.~\onlinecite{kotliar1999} which is related to the self-consistency
equation $F(\Gamma)=0$). We may, however, study the properties of the
Jacobian in the vicinity of the self-consistent solution
$\Gamma^\mathrm{sc}(\omega)$ by performing calculations with slightly
perturbed input hybridization functions:
\begin{equation}
\Gamma^\mathrm{input}(\omega) = \Gamma^\mathrm{sc}(\omega)
+ a \frac{e^{-b^2/4}}{b \sqrt{\pi}} e^{
-\left[ \ln(\omega/E)/b \right]^2 }.
\end{equation}
The perturbation takes the form of a log-Gaussian function centered at the
energy $E$ and of width $b$, similar to the commonly used broadening kernel
for producing smooth spectral functions in NRG (although the normalization
factor differs) \cite{bulla2008}. The weight $a$ should be chosen small
enough so that the response function
\begin{equation}
R_E(\omega) = \frac{1}{a} \left[ \Gamma^\mathrm{output}(\omega) -
\Gamma^{\mathrm{sc}}(\omega) \right]
\end{equation}
no longer depends on the value of $a$, but it must be large enough to
prevent numerical artifacts. The width $b$ should likewise be as small as
possible, although its value is ultimately limited by the NRG broadening
which restrains the energy resolution in $\Gamma^\mathrm{output}(\omega)$.
The calculations were performed for $a=0.001$ and $b=0.05$.

\begin{figure}[htbp]
\includegraphics[width=8cm,clip]{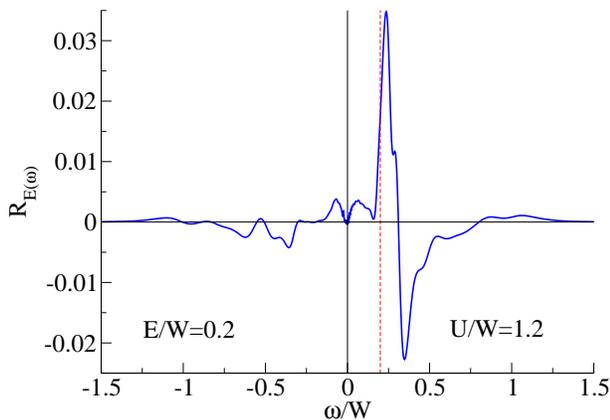}
\caption{(Color online) The response function $R_E(\omega)$ for the Hubbard
model for the excitation energy $E/W=0.2$ (graphically represented as the
vertical dashed line). Model parameters are the same as in Fig.~\ref{fig1}.}
\label{fig6}
\end{figure}

An example of the response function for the Hubbard model with
intermediately strong interaction $U/W=1.2$ is shown in Fig.~\ref{fig6}. It
reveals that a variation at a given excitation energy $E$ can lead to a
complex response at all energies. (However $R_E(\omega)$ vanishes in the
$\omega \to 0$ limit \cite{georges1992}.)
In simple linear mixing there is no exchange of information between
different energies (``cross-talk''), thus it takes many DMFT iterations for
reaching the self-consistency after a change has been imposed.
In the Broyden mixing, the application of the (approximate)
Jacobian effectively mixes the hybridization function at different energies,
thereby accelerating the propagation of the information.

\section{Unstable and metastable fixed points}
\label{sec8}

The concept of self-consistency is inseparably related to the concept of
iteration; this is directly implied by the form of the self-consistency
equation~\eqref{fixedpoint}. For this reason, the stability of the solutions
(fixed points) is related to the eigenspectrum of the DMFT transformation,
i.e. of the mapping $F$. Direct iteration can only be convergent if all the
eigenvalues $\lambda_i$ of the linearization of $F$ in the vicinity of the
fixed point are strictly less than 1 in absolute value, while it will
diverge when one or several eigenvalues are larger than one in absolute
value, unless the solution space is constrained in such a way that the
initial approximation for the solution has no components along the
directions of the corresponding eigenvectors. For linear mixing with
parameter $\alpha \in [0:1]$ (note that $\alpha=1$ corresponds to direct
iteration), the convergence criterium becomes \cite{dederichs1983}
\begin{equation}
\label{lin}
|1-\alpha (1-\lambda_i)| < 1.
\end{equation}
We denote by $\lambda_\mathrm{max}$ and $\lambda_\mathrm{min}$ the maximal
and minimal eigenvalue. If $\lambda_\mathrm{max} < 1$ and
$\lambda_\mathrm{min} > -1$, the direct iteration with $\alpha=1$ will
converge. If $\lambda_\mathrm{max} < 1$ and $\lambda_\mathrm{min} \leq -1$,
$\alpha$ should be $\alpha < 2/(1-\lambda_\mathrm{min})$. Finally, if
$\lambda_\mathrm{max} > 1$ the inequality~\eqref{lin} cannot be satisfied for
any $\alpha \in [0:1]$ and the linear mixing is of no help, thus the
use of advanced mixing schemes becomes mandatory.

\begin{figure}[htbp]
\includegraphics[width=8cm,clip]{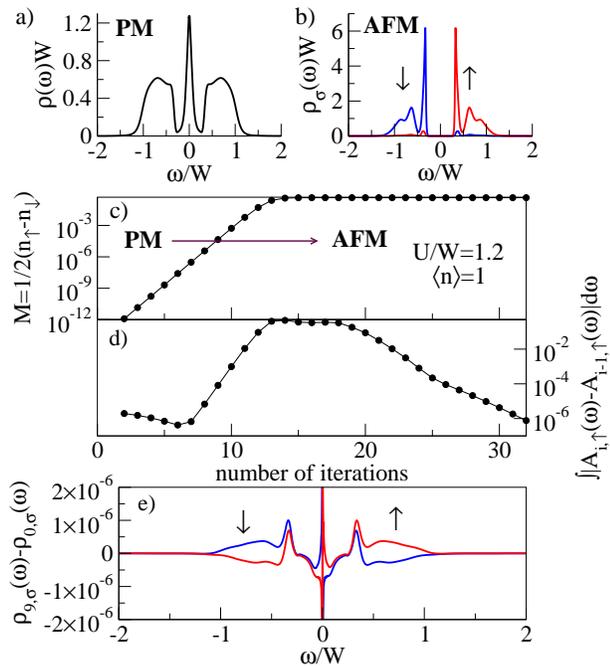}
\caption{(Color online) Evolution of the paramagnetic solution
for the half-filled Hubbard model after switching off the Broyden
mixing. a) Initial paramagnetic (PM) spectral function.
b) Resulting antiferromagnetic (AFM) spin-dependent spectral functions.
c) Magnetization and d) convergence as a function
of the number of iterations. e) Difference between the spectral functions
at iteration 9 and the initial spectral function indicating the progressive
breaking of the spin symmetry.
}
\label{fig7}
\end{figure}

As an example of a well-understood unstable solution, let us consider the
instability of the paramagnetic solution of the Hubbard model at
half-filling towards an antiferromagnetically ordered N\'eel ground state
\cite{jarrell1992, georges1992, zitzler2002, pruschke2005afm}. Using
Broyden's method, the paramagnetic (PM) solution can be stabilized in a
calculation which in principle allows a symmetry broken state. The system
drifts away, however, from the PM fixed point as soon as the Broyden mixing
is turned off and eventually it converges to an antiferromagnetic (AFM)
solution, as illustrated in Fig.~\ref{fig7}. The calculation was seeded with
a previously obtained self-consistent PM solution and iterated further
without Broyden solver. The magnetization immediately starts to increase
(Fig.~\ref{fig7}c) and by the tenth iteration the spectral functions develop
a narrow but sizeable singularity structure (emerging spectral gap) in the
quasiparticle peak, while the Hubbard bands start to become spin polarized
(Fig.~\ref{fig7}e). After 32 iterations, the result converged within
$\eta=10^{-6}$ to a stable self-consistent AFM solution shown in
Fig.~\ref{fig7}b.

It should be recalled that in a calculation where the $\mathrm{SU}(2)$
symmetry in spin space is explicitly maintained, the PM solution is stable
and the mapping $F$ is a contraction, as shown in Sec.~\ref{sec5}. The
eigenspectrum of the mapping $F$ is not only a property of the physical
model under consideration, but it also depends on the type of the long-range
order allowed for in the DMFT equations, and to some degree even on the
impurity solver used and on other details of the calculation (spectral
broadening, discretization parameter, number of states kept, etc.).

\begin{figure}[htbp]
\includegraphics[width=6cm,clip]{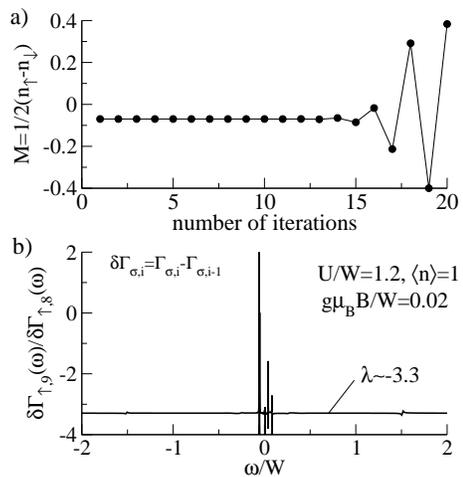}
\caption{Evolution of the solution for the Hubbard model at half-filling in
an external magnetic field after switching off the Broyden mixing. a)
Magnetization as a function of the number of iterations, b) ratio between
the differences of consecutive hybridization functions, which provides an
estimate for the dominant eigenvalue $\lambda \sim -3.3$ for $B/W=0.02$.}
\label{fig8}
\end{figure}

Another situation commonly encountered after switching off the Broyden
mixing is the emergence of oscillatory solutions which never converge. This
behavior can be observed for the Hubbard model at half-filling in a magnetic
field (see Sec.~\ref{sec6}). The polarized fixed-point solution are found to
be unstable and lead to oscillations between two almost fully spin-polarized
(in the opposite directions) spectral functions, see Fig.~\ref{fig8}a. The
instability can be traced to a dominant eigenvalue of $\lambda \sim -3.3 <
-1$ (for the example in Fig.~\ref{fig8}), as extracted from the ratio of
differences between consecutive hybridization functions in the vicinity of
the fixed point, Fig.~\ref{fig8}b. This solution could thus be stabilized
using linear mixing with $\alpha < 0.23$. It should be remarked that the
solution was found to be unstable for all magnetic fields that yield a
spin-polarized metallic solution, not only for weak fields where the system
is known to be unstable toward the AFM solution. If the instability of the
fixed point is a true physical instability also for large magnetic fields,
its nature is not very clear; it might correspond to canted ferromagnetism,
a tendency towards formation of spin density waves, or some other kind of
incommensurate order \cite{peters2007magnetic}. Since such states cannot be
described by the formalism used, the iteration cannot converge.

Since the fixed points $\Gamma^*$ of the self-consistency equation
$F(\Gamma^*)=0$ are generally stationary points, rather than extrema
\cite{kotliar1999, potthoff2003}, it will be interesting to further clarify
the relations between the stability of the DMFT iteration and the physical
stability of the solution, as well as their relation to the eigenspectra of
the mapping $F$ in the vicinity of the solutions. As demonstrated, the
proposed mixing technique can be a valuable tool for numerical studies of
these questions, since it allows in principle to obtain all self-consistent
solutions and (by turning the mixing off) to analyze the nature of their
possible instabilities.

\section{Conclusion}

It has been shown that the approach to the self-consistency can be greatly
accelerated by reformulating the DMFT loop as an iterative method for
solving a non-linear self-consistency equation using quasi-Newton-Raphson
methods. The tests performed for the paradigmatic case of the Hubbard model
at half-filling have shown that Broyden's method significantly outperforms
simple linear mixing. The approach is fully general and it can be also
applied when any other impurity solver (such as, for example, exact
diagonalisation, DMRG, or quantum Monte Carlo) is used; it appears likely
that similar speed-up factors could be achieved on equivalent problems. For
particularly pathological situations, the improvement might be sufficient to
bring previously forbidding problems within reach. This is particularly
important near quantum phase transitions, where reaching the convergence
becomes problematic due to critical slowing down and the detailed behavior
at the transition points is still a matter of controversy for many important
problems. The acceleration due to the use of Broyden's method might be
instrumental in answering some of these long-standing questions.  In
addition, the solver can be used to stabilize unstable fixed-point solutions
and to study their properties. Since the solver is robust, easy to implement
and to incorporate in the DMFT cycle, there is little reason not to use it.

\begin{acknowledgments}
Very fruitful discussions with Thomas Pruschke and Robert Peters are
gratefully acknowledged.
\end{acknowledgments}

\bibliography{paper}

\end{document}